\documentclass[letter]{aa}
\usepackage{graphicx}
\usepackage[varg]{txfonts}
\usepackage{natbib}
\usepackage{adjmulticol}
\usepackage{lipsum}
\bibpunct{(}{)}{;}{a}{}{,}
\usepackage{multicol,lipsum}
\usepackage{soul}

\begin{document} 

\title{Multiwavelength Galactic center gamma-ray observations explained by a unified cosmic-ray dynamics model}
\subtitle{}

\titlerunning{Galactic Center gamma-rays explained by cosmic rays}

\author{Andrés Scherer\inst{1,2}
\and Jorge Cuadra\inst{3,4}}

\institute{Departamento de Física, Universidad de Santiago de Chile (USACH), Av. Victor Jara 3493, Estación Central, Santiago, Chile
\and Center for Interdisciplinary Research in Astrophysics and Space Sciences (CIRAS), Universidad de Santiago de Chile, Chile
\and Departamento de Ciencias, Facultad de Artes Liberales, Universidad Adolfo Ibáñez, Av. Padre Hurtado 750, Viña del Mar, Chile
\and Millennium Nucleus on Transversal Research and Technology to Explore Supermassive Black Holes (TITANS), Chile}

\date{Received 26 July 2024 / Accepted 26 September 2024}

\abstract
{High-energy (HE) and very high-energy (VHE) gamma-ray observations from the Galactic center (GC) detected extended emission correlated with the morphology of the central molecular zone (CMZ). Emission in both bands is expected to be produced by hadronic interaction between cosmic rays (CRs) and ambient gas.}
{We examine if our three previously proposed scenarios for the CR sources and dynamics, which are consistent with the VHE gamma-ray data (1–100 TeV), also match the HE gamma-ray observations (10–300 GeV). Additionally, we analyze the effect of the isotropic Galactic CR "sea" inside the CMZ.}{We generated synthetic gamma-ray maps considering a simplified isotropic diffusion, but more realistic dynamics with two diffusion zones (in and out of the CMZ) and polar advection, for mono-energetic particles of 3 TeV. Additionally, we considered two gas distributions for the CMZ (with and without an inner cavity), and CR populations injected from the clusters of young massive stars (the Arches Cluster, the Quintuplet Cluster, and the nuclear star cluster), plus the supernova Sgr A East.}{Only the combination of more realistic CR dynamics, the CMZ with an inner cavity, CR injection from all proposed sources, and a CR sea similar to that observed in the Solar System reproduced the current HE and VHE gamma-ray detection from the CMZ and was consistent with the observed gamma-rays from Sagittarius A*.}{The HE and VHE gamma-rays observations of the GC can be reproduced by a unified model for the CRs.} 


\maketitle

\section{Introduction}

Very high-energy (VHE) gamma-ray observations (1--100 TeV) from the Galactic center (GC) by the High Energy Stereoscopic System \citep[HESS;][]{Aharonian.2009,HESS.2016,HESS.2018}, the Major Atmospheric Gamma-Ray Imaging Cherenkov \citep[MAGIC;][]{MAGIC.2020}, and the Very Energetic Radiation Imaging Telescope Array System \citep[Veritas;][]{VERITAS.2021} all detected an extended emission correlated with the morphology of the high-density molecular clouds in the center of our galaxy, an area known as the central molecular zone \citep[CMZ;][]{Battersby.2020,Henshaw.2023}. This emission is typically explained by the proton--proton collisions between cosmic rays (CRs) and the ambient gas, which produce neutral pi-mesons ($\pi^0$) that decay into observable gamma rays. Assuming this mechanism, the inferred energy density profile and spectrum of CRs suggest a central and continuous source of protons with energies up to 1 PeV \citep{HESS.2016}. On the other hand, from the analysis of high-energy (HE) gamma rays (10–300 GeV) measured by the Fermi Gamma-ray Space Telescope (Fermi Lat), \citet{Gaggero.2017} and \citet{Huang.2021} report an extended emission coherent with the CMZ, which is consistent with the same kind of hadronic interaction.

Assuming that the gamma rays in both bands are produced by a common CR population, Galactic sources of protons accelerated up to 1 PeV are required, and their nature is still under debate \citep{Vieu.2022, Vieu.2023}. Nevertheless, several sources have been proposed. Protons could be accelerated in the vicinity of the supermassive black hole Sagittarius A* \citep[Sgr A*;][]{HESS.2016,Muena.2024} or within the Arches Cluster (AC), the Quintuplet Cluster (QC), and the nuclear star cluster (NSC) of young massive stars \citep{Aharonian.2019}. In addition to these continuous sources, the supernova Sagittarius A East \citep[SN Sgr A East;][]{Ekers.1983,Maeda.2002} could have generated an impulsive contribution of accelerated particles. These ideas were tested in our previous works.  We showed that if simplified CR dynamics are considered (i.e., only isotropic diffusion with a constant coefficient), CRs need to be accelerated from the NSC for a disk-like gas distribution of the CMZ, or alternatively from the NSC and from SN Sgr A East for a ring-like gas distribution \citep[][hereafter Paper~I]{Scherer.2022}. However, when more realistic CR dynamics are considered (i.e., low diffusion within the CMZ and regular Galactic-disk diffusion outside the CMZ, plus polar advection), the CRs need to be accelerated from several sources, namely the AC, QC, and NSC, plus SN Sgr A East, and the gas must have a ring-like distribution \citep[][hereafter Paper~II]{Scherer.2023}.

Outside of the GC, CR transport in the Galactic disk is known to be dominated by diffusion, which produces an isotropic Galactic CR population of particles injected from several sources, the so-called CR sea. Due to the transport nature, the CR sea can only be detected directly in the Solar System \citep{Blasi.2013,Gaisser.2016}, and its contribution in the CMZ is debated. Gamma-ray observations and CR models are expected to constrain it, but both an overdensity of the CR sea generated by the radial dependence of the diffusion toward the GC \citep{Gaggero.2017} and a CR suppression due to a barrier created by the strong magnetic field \citep{Huang.2021} have been invoked in different models to explain the HE gamma-ray observations by Fermi LAT.

In Paper~I we explored the influence of the three-dimensional (3D) shape of the CMZ on the indirect observation of the CR energy density via gamma-ray detection, and in Paper~II our goal was to verify if more realistic CR dynamics for the GC environment are consistent with current gamma-ray observations. From both, we determined three different scenarios consistent with VHE gamma-ray observations \citep{HESS.2016}. In this paper, we test whether and which of those satisfactory scenarios also match the HE gamma-ray detection by Fermi LAT \citep{Gaggero.2017}, and study the CR sea contribution inside the CMZ. Our aim is to propose a unified scenario for both gamma-ray bands.

\section{Methodology} \label{sec_metodo} 

We computed the CR dynamics as in Papers I and II for less energetic particles and determined whether if the models found to reproduce the HESS observations are also consistent with Fermi LAT detections. In Papers I and II we computed the CR dynamics, modeled the CMZ morphology, and calculated the gamma-ray luminosity in a 3D grid in order to find which models match the gamma-ray observation. More details are available in Sect.~2 of Papers I and II, which are summarized in appendix \ref{App_model} here.

In Papers I and II, we neglected the VHE gamma-ray emitted by the CR sea because it is considered to be relatively weak in the GC at that energy range \citep{HESS.2018}. However, it is not clear if the same is true for the HE gamma-ray band. Therefore, we included it in some of our current models as a static component. Its energy density was assumed to have a homogeneous value $w_{CR}$ = 0.036 eV cm$^{-3}$, which was obtained by integrating the CR spectrum between 0.1 and 3 TeV for particles observed directly in the Solar System \citep[i.e., $I(E) \approx 1.7 \times 10^4$(E/GeV)$^{-2.7}$ nucleons m$^{-2}$ s$^{-1}$ sr$^{-1}$ GeV$^{-1}$;][]{Aharonian.2004,Longair.2011,Gaisser.2016}.

\section{Results}

We only used the models that were found in Papers I and II to satisfactorily reproduce current VHE gamma-ray observations by HESS. Those models are: (i) simplified CR transport, injection from the NSC, and the CMZ as a disk (NSC/CMZ-disk); (ii) simplified CR transport, injection from the NSC and SN Sgr A East, and the CMZ as a ring (NSC+SgrAE/CMZ-ring); and (iii) more realistic CR transport, injection from the NSC, AC, QC and SN Sgr A East, and the CMZ as a ring (All/CMZ-ring). These three options were considered with and without the contribution of the CR sea within the CMZ. 

\subsection{Model results}\label{model_results}

\begin{figure*}
\centering
\includegraphics[width=17cm]{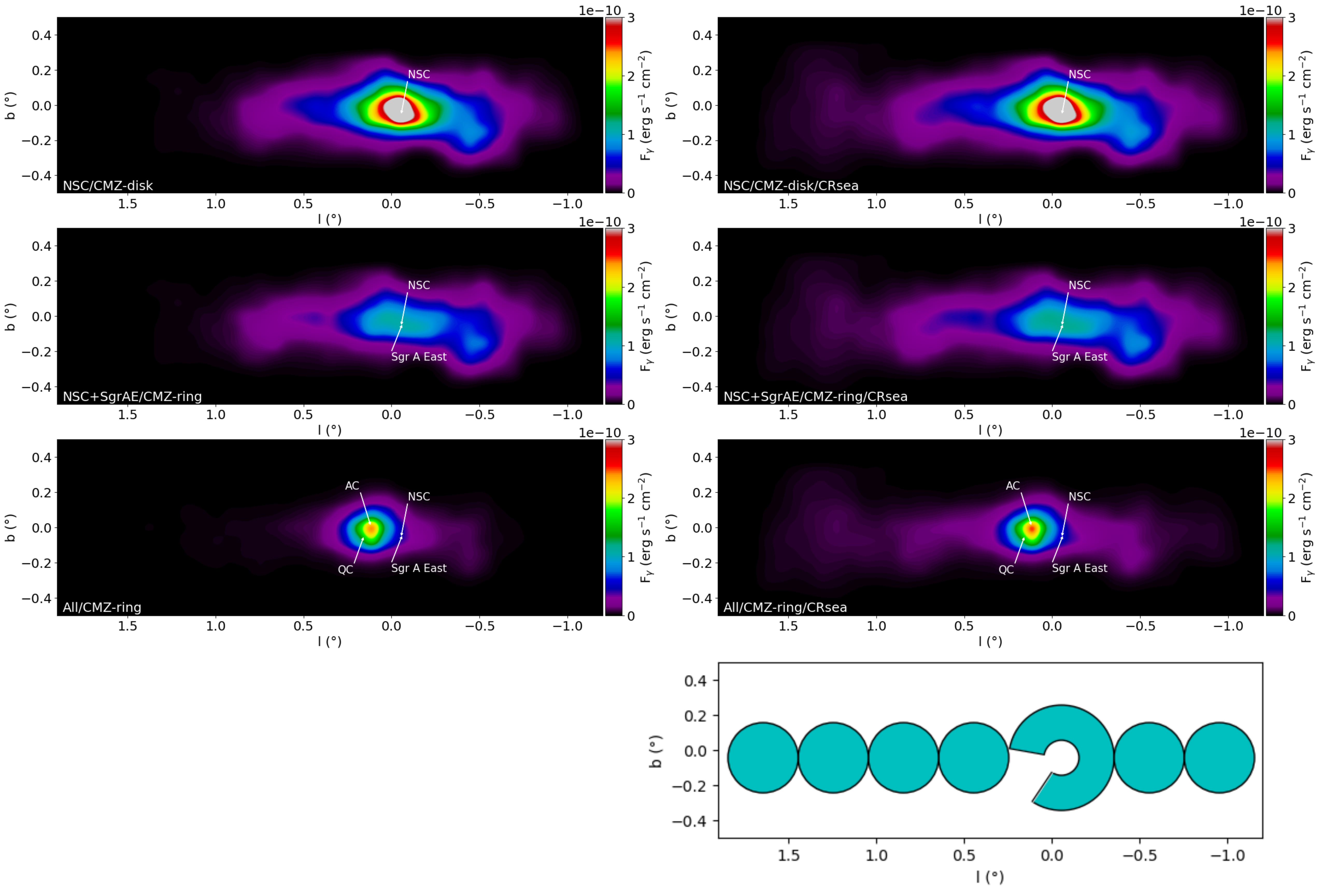}
\caption{HE gamma-ray synthetic maps computed as observed from Earth for models consistent with VHE observations. Top and second row: Synthetic maps for simplified CR dynamics with (left) and without (right) the CR sea. Third row: Synthetic maps for more realistic CR dynamics with (left) and without (right) the CR sea. The gas geometry and CR sources are indicated in each panel. All maps have been smoothed with a 0.08$\degr$ Gaussian function to approximately adopt the best Fermi Lat beamwidth. Bottom-right panel: Regions where we contrasted our models with Fermi Lat observations \citep{Gaggero.2017}.}
\label{fig_models}
\end{figure*}

The left panels of Fig.~\ref{fig_models} show the resulting HE gamma-ray flux maps without the CR sea for NSC/CMZ-disk (top panel), NSC+SgrAE/CMZ-ring (second panel), and All/CMZ-ring (third panel). The top three panels on the right show the same models but including the CR sea as detected in the Solar System (model tags with the suffix /CRsea). All maps have been smoothed with a 0.08$\degr$ Gaussian function to adjust to the best Fermi Lat resolution.

The bottom-right panel in Fig.~\ref{fig_models} shows the regions used to contrast our models with Fermi Lat observations, following \cite{Gaggero.2017}. We integrated the gamma-ray flux over those regions and obtained gamma-ray luminosities (L$_\gamma$). These areas correspond to an annulus centered in Sgr A* (l=-0.056$\degr$, b=-0.04588$\degr$) with inner and outer radii of 0.1$\degr$ and 0.3$\degr$, but excluding the region between +10$\degr$ and -56$\degr$ from the positive Galactic longitude axis, and six circular areas with a radius of 0.2$\degr$ centered on b=-0.04588$\degr$ and l=-0.956$\degr$, -0.556$\degr$, 0.444$\degr$, 0.844$\degr$, 1.244$\degr$, and 1.644$\degr$. We took the L$_\gamma$ measured by Fermi Lat from \citet[]{Gaggero.2017}, using their Eq. 3, the values from their Fig.~3, and the column density obtained from the CS line by \cite{Tsuboi.1999}\footnote{Available at \url{https://www.nro.nao.ac.jp/~nro45mrt/html/results/data.html}.  Notice that in our model we trace the gas mass with the CO line (see appendix \ref{subsub-3D-CMZ}), but here we use CS to recover the L$_\gamma$ values from \citet{Gaggero.2017}.}. 

\begin{figure*}
\centering
\includegraphics[width=17cm]{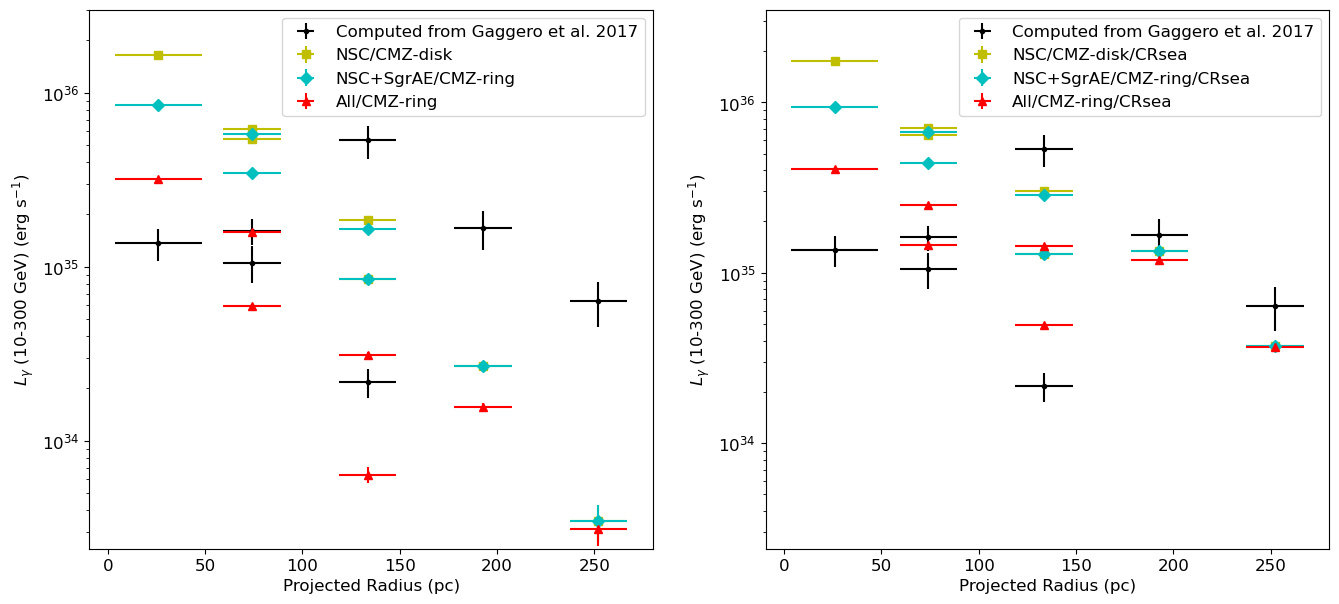}
\caption{CMZ gamma-ray luminosity profiles in the HE band from models consistent with VHE observations, with and without the CR sea (right and left panel, respectively). The luminosities were calculated by integrating the flux in the regions shown in the bottom-right panel of Fig. \ref{fig_models}. Black crosses denote the observed gamma-ray luminosity from \citet{Gaggero.2017} in the CMZ. Yellow squares represent the gamma-ray luminosity for simplified CR transport, injection from the NSC, and the CMZ as a disk (NSC/CMZ-disk). Cyan diamonds represent the gamma-ray luminosity for simplified CR transport, injection from the NSC and SN Sgr A East, and the CMZ as a ring (NSC+SgrAE/CMZ-ring). Red triangles represent the gamma-ray luminosity for more realistic CR transport, injection from the NSC, AC, QC and SN Sgr A East, and the CMZ as a ring (All/CMZ-ring)}
\label{fig_lum_all}
\end{figure*}

Figure~\ref{fig_lum_all} shows the comparison between the observed L$_\gamma$ and that obtained from our models with and without the CR sea (right and left panel respectively). Luminosities are shown as a function of the projected radius from Sgr A*; therefore, two data points are plotted at $\approx$75 pc and $\approx$135 pc due to the regions selected. It is clear that only the models that consider the gamma-ray production by the CR sea satisfactorily reproduce the observations at $R>100$ pc, while models without the CR sea are significantly steeper. For $R<100$, the All/CMZ-ring/CRsea model most closely matches the Fermi Lat measurements.

\begin{figure}
\centering
\resizebox{\hsize}{!}{\includegraphics{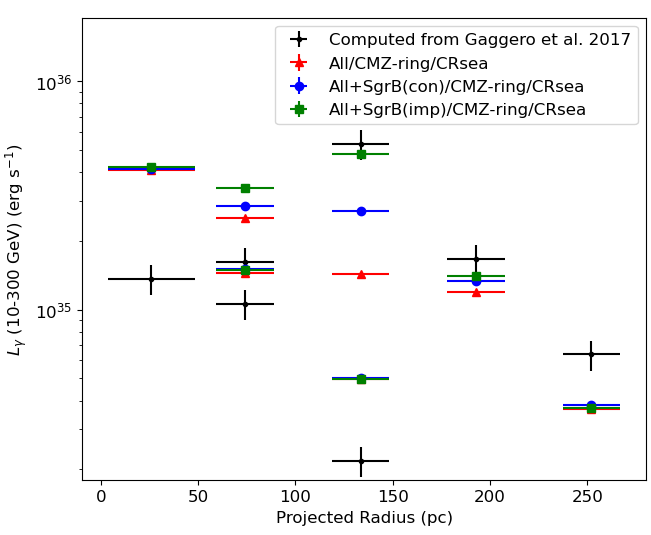}}
\caption{Same as the right panel of Fig. \ref{fig_lum_all}, but now plotting the gamma-ray luminosity for the All/CMZ-ring/CRsea model (red triangles), All/CMZ-ring/CRsea plus Sgr B as an impulsive source (blue circles), and All/CMZ-ring/CRsea plus Sgr B as a continuous source (green square).}
\label{fig_lum_all+sgrB}
\end{figure}

There are noticeable mismatches at $R \approx 25$ and 135 pc.
The emission from the inner region is discussed in Sect.~\ref{conclusions}.
To explain the high flux measured at $R \approx 135$ pc, we performed additional tests, including a possible source whose properties are not well constrained. This was motivated by the observations of the star formation complex Sagittarius B \citep[Sgr B;][]{Ginsburg.2018,Henshaw.2023} by Fermi LAT \citep{Yang.2015}. This emission is not observed individually by HESS due to its spectral index and flux. We studied the effect of an impulsive or continuous CR injection at this location in the sky, at the line-of-sight coordinate from \citet{Kruijssen.2015}. Following the methodology described in Sect.~\ref{sec_metodo} and considering that Sgr B's stellar mass \citep[$\sim 4 \times 10^4$ M$_\sun$;][]{Barnes.2017} is similar to that of the AC \citep[$\sim 4 \times 10^4$ M$_\sun$;][]{Clarkson.2012}, we computed the gamma-ray emission produced by a continuous CR source to be $A_0=3.2 \times 10^{36}$ erg s$^{-1}$ over $10^6$ yr (see Sect.~\ref{subsub_source}). Next, we modeled an impulsive source as a low-energy supernova ($B_0=1.2\times 10^{49}$ erg) that exploded $10^4$ yr ago. In Fig. \ref{fig_lum_all+sgrB}, we compare both simulations added to the All/CMZ-ring/CRsea model shown in Fig. \ref{fig_lum_all}. Including either the continuous or impulsive source for Sgr B improves our All/CMZ-ring/CRsea model at $R \approx 135$ pc.

\begin{figure}
\resizebox{\hsize}{!}{\includegraphics{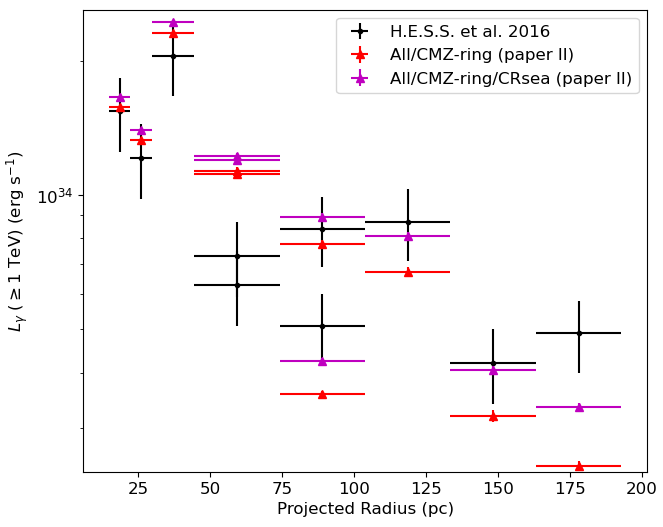}}
\caption{Gamma-ray luminosity profiles from Sgr A* along the Galactic disk for the All/CMZ-ring models in the VHE gamma-ray band. Red triangles show our Paper~II results, which neglected the CR sea, while the magenta triangles include it.}
\label{fig_hess+sea}
\end{figure}

In Papers I and II we neglected the CR sea, as it was not expected to be significant for VHE gamma rays. In this study we find that the VHE profiles are somewhat affected by its inclusion at $R>75$pc, and that its inclusion actually produces a better match to the data (see Fig. \ref{fig_hess+sea}). In the same way as described in Sect.~\ref{sec_metodo}, we computed the CR sea between 0.01 and 1 PeV, which is 0.001 eV cm$^{-3}$. Therefore, our All/CMZ-ring/CRsea model provides a common scenario that can be used to interpret the gamma-ray observations by both HESS and Fermi LAT.

\subsection{Gamma rays from Sgr A*} \label{gammaSgrA}

In Paper~II we determined that our models are consistent with an additional gamma-ray source at the location of Sgr A*. 
This additional emission could explain its spectral mismatch with the CMZ, namely the spectral index observed by \citet{Chernyakova.2011} for Sgr A* is $\Gamma=2.68 \pm 0.05$, while the one reported by \citet{Gaggero.2017} for the CMZ is $\Gamma=2.41^{+0.07}_{-0.06}$.
Here, we compared our flux prediction in a circular region of radius 0.1$\degr$ centered at Sgr A* with that reported by \citet{Chernyakova.2011}. We integrated their spectrum between the relevant gamma-ray energies and find that the observed flux is 2.5 times lower than that predicted by our NSC/CMZ-disk/CRsea model, is similar to that predicted by our NSC+SgrAE/CMZ-ring/CRsea model, and is 2.1 times higher than that predicted by our All/CMZ-ring/CRsea model. 
Regarding the gamma-ray source added to address the spectral inconsistencies, only the All/CMZ-ring/CRsea model is coherent with HE and VHE gamma-ray observation from Sgr A*, in agreement with the conclusion of Sect.~\ref{model_results}. Finally, as described in Paper~II, including a new CR source to produce the additional Sgr A* gamma-ray emission strongly modifies the profiles shown in Fig. \ref{fig_lum_all}, making them inconsistent.

\subsection{Gamma-ray spectra} \label{subsub_spect}

\begin{figure}
\resizebox{\hsize}{!}{\includegraphics{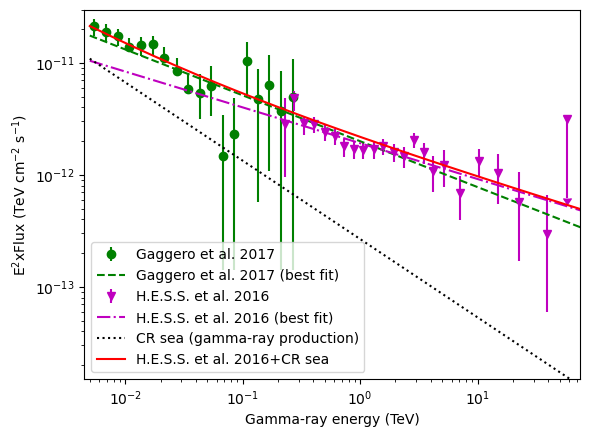}}
\caption{Gamma-ray spectrum observed by HESS and Fermi LAT in the CMZ. Magenta triangles are the VHE gamma rays reported by \citet{HESS.2016} and green circles are the HE gamma rays reported by \citet{Gaggero.2017}. The dot-dash magenta line is the best fit by \citet{HESS.2016} considering only HESS data, and the dashed green line is the best fit by \citet{Gaggero.2017} considering both observations. The dotted black line is the CR sea observed in the Solar System. The solid red line is the HESS best fit plus the CR sea.}
\label{fig_spec_cmz}
\end{figure}

Our results indicate that there is a unique scenario that explains observations across the HE and VHE gamma-ray bands. Therefore the spectra detected by Fermi LAT and HESS must be connected. 
Figure~\ref{fig_spec_cmz} shows the HESS and Fermi LAT data and the best power-law fit for each dataset extracted from \citet{HESS.2016} and \citet{Gaggero.2017}, respectively. The HESS power law does not fit the tight error bars for the flux at $\sim10$ GeV. However, if we add the expected contribution from the CR sea within the CMZ, the two datasets are well matched. We computed the CR sea spectrum from Eq. \ref{eq_lum} considering the spectral index detected in the Solar System (see Sect. \ref{sec_metodo}). We can therefore conclude that the whole VHE and HE spectral range is produced by the same composite CR population. We note that the CR sea is negligible at VHE energies; therefore, the observed spectrum by HESS is produced primarily by the CRs injected by sources located in the GC. However, as the CR sea is important at lower energies, for the Fermi Lat detection both the CR sea and the GC sources are relevant.   

In appendix \ref{App_SgrA*_spec} we repeat the same process for Sgr A*, finding that it is not possible to connect the \citet{HESS.2016} spectrum with that observed by Fermi-LAT \citep{Chernyakova.2011}, not even if we add five times the CR sea observed in the Solar System. This reinforces our conclusion from Sect.~\ref{gammaSgrA}, that an additional central gamma-ray source is required. 

\section{Discussion and conclusions} \label{conclusions}

In Paper~I we studied the dependence of the gamma-ray flux on the 3D gas distribution of the CMZ using a simple recipe for CR diffusion. In Paper~II we verified that more realistic CR dynamics can also produce results consistent with the current gamma-ray observations. In fact, we found three satisfactory models that combine different ingredients to reproduce the HESS data, and suggested that the upcoming Cherenkov Telescope Array \citep[CTA;][]{CTA.2019}, thanks to its higher resolution and sensitivity, will break the degeneracy and allow us to identify the CR sources, or "PeVatrons", in the GC. 

Here, we took an alternative approach and computed the HE gamma rays produced by these three models to contrast them with existing Fermi LAT observations. Our results show that the more realistic CR dynamics with two diffusion zones and polar advection, the CMZ as a ring, a CR population composed of four sources, and a CR sea similar to that of the Galactic disk are required to reproduce the observations in both gamma-ray bands. This unified scenario is based on well-known physical processes and sources, with fiducial values taken from the literature or observations at other wavelengths, as detailed in Paper~II.  An additional source, associated with Sgr B but without well-constrained properties, improves the fit even further. 
Nevertheless, we acknowledge that the values we used are still uncertain, and that other models could also reproduce the data. For example, \citet{Dorner.2024} used an anisotropic CR transport prescription based on the \citet{Guenduez.2020} magnetic field configuration in the GC, with CRs  accelerated from Sgr A*, the supernova remnant G0.9+01, and the pulsar candidate HESS J1746-285. Moreover, the 3D morphology of the CMZ based on radio observations is still under debate \citep{Sofue.2022,Henshaw.2023}, and it might be inconsistent with our results.
Because of all this, CTA data will still be required to better understand this region.

All our models overestimate the gamma-ray flux at $R \approx 25$ (see Fig.~\ref{fig_lum_all}), which is not an issue for VHE gamma rays (see Fig. \ref{fig_hess+sea}). This discrepancy could be related to our chosen polar advection velocity, a physical mechanism that is more relevant for less energetic CRs given their slower diffusion (see Sect. \ref{Cr_dim}). According to \citet{Guo.2012}, the Fermi bubbles' outflow velocity could be $\approx 30$ times higher than the 1000 km s$^{-1}$ we used, taken from \citet{Bordoloi.2017}.  Such a strong advection would quickly deplete the particles producing the HE gamma-ray flux without significantly modifying the dynamics of the CRs that produce the VHE emission.

Cosmic rays from outside the GC can also be relevant in this region. The magnitude of the CR sea inside the CMZ is not clear yet, and both CR suppression \citep{Huang.2021} and CR overdensity \citep{Gaggero.2017} have been proposed. Our results show an intermediate scenario, in which the CR sea at the CMZ has the same level as observed in the Solar System. 

Furthermore, we  have strengthened our conclusion from Paper~II that our model requires an additional point-like gamma-ray source at the location of Sgr A*. This emission is likely be produced within the CMZ cavity by a physical process, gas structure, or CR population that does not interact with nor is part of the CMZ, as propounded by \citet{Aharonian.2005} and \citet{Chernyakova.2011}. Another possibility proposed recently by \citet{Muena.2024} is that Sgr A* shows an energy-dependent transition due to nonuniform diffusion and advection in the central parsec, which generates a broken power-law spectrum.

\begin{acknowledgements}
We thank the anonymous referee for their useful comments. This project was partially funded by the Max Planck Society through a “Partner Group” grant. AS acknowledges financial support from POSTDOC\_DICYT 042431ER\_Postdoc Vicerrectoría de Investigación, Innovación y Creación (USACH). AS and JC acknowledge financial support from ANID, FONDECYT Regular 1211429. JC acknowledges ANID further through Millennium Science Initiative Program NCN$2023\_002$. The Geryon cluster at the Centro de Astro-Ingenieria UC was extensively used for the calculations performed in this paper. BASAL CATA PFB-06, the Anillo ACT-86, FONDEQUIP AIC-57, and QUIMAL 130008 provided funding for several improvements to the Geryon cluster.
\end{acknowledgements}

\bibliographystyle{aa}
\bibliography{biblio}

\begin{thebibliography}{53}
\expandafter\ifx\csname natexlab\endcsname\relax\def\natexlab#1{#1}\fi

\bibitem[{{Abeysekara} {et~al.}(2017){Abeysekara}, {Albert}, {Alfaro},
  {Alvarez}, {{\'A}lvarez}, {Arceo}, {Arteaga-Vel{\'a}zquez}, {Avila Rojas},
  {Ayala Solares}, {Barber}, {Bautista-Elivar}, {Becerril}, {Belmont-Moreno},
  {BenZvi}, {Berley}, {Bernal}, {Braun}, {Brisbois}, {Caballero-Mora},
  {Capistr{\'a}n}, {Carrami{\~n}ana}, {Casanova}, {Castillo}, {Cotti},
  {Cotzomi}, {Couti{\~n}o de Le{\'o}n}, {De Le{\'o}n}, {De la Fuente},
  {Dingus}, {DuVernois}, {D{\'\i}az-V{\'e}lez}, {Ellsworth}, {Engel},
  {Enr{\'\i}quez-Rivera}, {Fiorino}, {Fraija}, {Garc{\'\i}a-Gonz{\'a}lez},
  {Garfias}, {Gerhardt}, {Gonz{\'a}lez Mu{\~n}oz}, {Gonz{\'a}lez}, {Goodman},
  {Hampel-Arias}, {Harding}, {Hern{\'a}ndez}, {Hern{\'a}ndez-Almada}, {Hinton},
  {Hona}, {Hui}, {H{\"u}ntemeyer}, {Iriarte}, {Jardin-Blicq}, {Joshi},
  {Kaufmann}, {Kieda}, {Lara}, {Lauer}, {Lee}, {Lennarz}, {Vargas},
  {Linnemann}, {Longinotti}, {Luis Raya}, {Luna-Garc{\'\i}a}, {L{\'o}pez-Coto},
  {Malone}, {Marinelli}, {Martinez}, {Martinez-Castellanos},
  {Mart{\'\i}nez-Castro}, {Mart{\'\i}nez-Huerta}, {Matthews},
  {Miranda-Romagnoli}, {Moreno}, {Mostaf{\'a}}, {Nellen}, {Newbold}, {Nisa},
  {Noriega-Papaqui}, {Pelayo}, {Pretz}, {P{\'e}rez-P{\'e}rez}, {Ren}, {Rho},
  {Rivi{\`e}re}, {Rosa-Gonz{\'a}lez}, {Rosenberg}, {Ruiz-Velasco}, {Salazar},
  {Salesa Greus}, {Sandoval}, {Schneider}, {Schoorlemmer}, {Sinnis}, {Smith},
  {Springer}, {Surajbali}, {Taboada}, {Tibolla}, {Tollefson}, {Torres},
  {Ukwatta}, {Vianello}, {Weisgarber}, {Westerhoff}, {Wisher}, {Wood},
  {Yapici}, {Yodh}, {Younk}, {Zepeda}, {Zhou}, {Guo}, {Hahn}, {Li}, \&
  {Zhang}}]{Abeysekara.2017}
{Abeysekara}, A.~U., {Albert}, A., {Alfaro}, R., {et~al.} 2017, Science, 358,
  911

\bibitem[{{Ackermann} {et~al.}(2014){Ackermann}, {Albert}, {Atwood}, {Baldini},
  {Ballet}, {Barbiellini}, {Bastieri}, {Bellazzini}, {Bissaldi}, {Blandford},
  {Bloom}, {Bottacini}, {Brandt}, {Bregeon}, {Bruel}, {Buehler}, {Buson},
  {Caliandro}, {Cameron}, {Caragiulo}, {Caraveo}, {Cavazzuti}, {Cecchi},
  {Charles}, {Chekhtman}, {Chiang}, {Chiaro}, {Ciprini}, {Claus},
  {Cohen-Tanugi}, {Conrad}, {Cutini}, {D'Ammando}, {de Angelis}, {de Palma},
  {Dermer}, {Digel}, {Di Venere}, {Silva}, {Drell}, {Favuzzi}, {Ferrara},
  {Focke}, {Franckowiak}, {Fukazawa}, {Funk}, {Fusco}, {Gargano}, {Gasparrini},
  {Germani}, {Giglietto}, {Giordano}, {Giroletti}, {Godfrey}, {Gomez-Vargas},
  {Grenier}, {Guiriec}, {Hadasch}, {Harding}, {Hays}, {Hewitt}, {Hou},
  {Jogler}, {J{\'o}hannesson}, {Johnson}, {Johnson}, {Kamae}, {Kataoka},
  {Kn{\"o}dlseder}, {Kocevski}, {Kuss}, {Larsson}, {Latronico}, {Longo},
  {Loparco}, {Lovellette}, {Lubrano}, {Malyshev}, {Manfreda}, {Massaro},
  {Mayer}, {Mazziotta}, {McEnery}, {Michelson}, {Mitthumsiri}, {Mizuno},
  {Monzani}, {Morselli}, {Moskalenko}, {Murgia}, {Nemmen}, {Nuss}, {Ohsugi},
  {Omodei}, {Orienti}, {Orlando}, {Ormes}, {Paneque}, {Panetta}, {Perkins},
  {Pesce-Rollins}, {Petrosian}, {Piron}, {Pivato}, {Rain{\`o}}, {Rando},
  {Razzano}, {Razzaque}, {Reimer}, {Reimer}, {S{\'a}nchez-Conde}, {Schaal},
  {Schulz}, {Sgr{\`o}}, {Siskind}, {Spandre}, {Spinelli}, {Stawarz}, {Strong},
  {Suson}, {Tahara}, {Takahashi}, {Thayer}, {Tibaldo}, {Tinivella}, {Torres},
  {Tosti}, {Troja}, {Uchiyama}, {Vianello}, {Werner}, {Winer}, {Wood}, {Wood},
  \& {Zaharijas}}]{Ackermann.2014}
{Ackermann}, M., {Albert}, A., {Atwood}, W.~B., {et~al.} 2014, \apj, 793, 64

\bibitem[{{Adams} {et~al.}(2021){Adams}, {Benbow}, {Brill}, {Brose},
  {Buchovecky}, {Capasso}, {Christiansen}, {Chromey}, {Daniel}, {Errando},
  {Falcone}, {Feng}, {Finley}, {Fortson}, {Furniss}, {Gent}, {Gillanders},
  {Giuri}, {Hanna}, {Hervet}, {Holder}, {Hughes}, {Humensky}, {Jin}, {Kaaret},
  {Kelley-Hoskins}, {Kertzman}, {Kieda}, {Krennrich}, {Kumar}, {Lang}, {Lundy},
  {Maier}, {Moriarty}, {Mukherjee}, {Nieto}, {Nievas-Rosillo}, {O'Brien},
  {Ong}, {Otte}, {Pfrang}, {Pohl}, {Prado}, {Pueschel}, {Quinn}, {Ragan},
  {Reynolds}, {Ribeiro}, {Richards}, {Roache}, {Ryan}, {Santander},
  {Schlenstedt}, {Sembroski}, {Shang}, {Stevenson}, {Wakely}, {Weinstein}, \&
  {Williams}}]{VERITAS.2021}
{Adams}, C.~B., {Benbow}, W., {Brill}, A., {et~al.} 2021, \apj, 913, 115

\bibitem[{{Aharonian} {et~al.}(2009){Aharonian}, {Akhperjanian}, {Anton},
  {Barres de Almeida}, {Bazer-Bachi}, {Becherini}, {Behera}, {Bernl{\"o}hr},
  {Boisson}, {Bochow}, {Borrel}, {Braun}, {Brion}, {Brucker}, {Brun},
  {B{\"u}hler}, {Bulik}, {B{\"u}sching}, {Boutelier}, {Chadwick},
  {Charbonnier}, {Chaves}, {Cheesebrough}, {Chounet}, {Clapson}, {Coignet},
  {Dalton}, {Daniel}, {Davids}, {Degrange}, {Deil}, {Dickinson},
  {Djannati-Ata{\"\i}}, {Domainko}, {O'C. Drury}, {Dubois}, {Dubus}, {Dyks},
  {Dyrda}, {Egberts}, {Emmanoulopoulos}, {Espigat}, {Farnier}, {Feinstein},
  {Fiasson}, {F{\"o}rster}, {Fontaine}, {F{\"u}{\ss}ling}, {Gabici}, {Gallant},
  {G{\'e}rard}, {Giebels}, {Glicenstein}, {Gl{\"u}ck}, {Goret}, {Hauser},
  {Hauser}, {Heinz}, {Heinzelmann}, {Henri}, {Hermann}, {Hinton}, {Hoffmann},
  {Hofmann}, {Holleran}, {Hoppe}, {Horns}, {Jacholkowska}, {de Jager}, {Jung},
  {Katarzy{\'n}ski}, {Katz}, {Kaufmann}, {Kendziorra}, {Kerschhaggl},
  {Khangulyan}, {Kh{\'e}lifi}, {Keogh}, {Komin}, {Kosack}, {Lamanna}, {Lenain},
  {Lohse}, {Marandon}, {Martin}, {Martineau-Huynh}, {Marcowith}, {Maurin},
  {McComb}, {Medina}, {Moderski}, {Moulin}, {Naumann-Godo}, {de Naurois},
  {Nedbal}, {Nekrassov}, {Niemiec}, {Nolan}, {Ohm}, {Olive}, {de O{\~n}a
  Wilhelmi}, {Orford}, {Ostrowski}, {Panter}, {Paz Arribas}, {Pedaletti},
  {Pelletier}, {Petrucci}, {Pita}, {P{\"u}hlhofer}, {Punch}, {Quirrenbach},
  {Raubenheimer}, {Raue}, {Rayner}, {Renaud}, {Rieger}, {Ripken}, {Rob},
  {Rolland}, {Rosier-Lees}, {Rowell}, {Rudak}, {Rulten}, {Ruppel}, {Sahakian},
  {Santangelo}, {Schlickeiser}, {Sch{\"o}ck}, {Schr{\"o}der}, {Schwanke},
  {Schwarzburg}, {Schwemmer}, {Shalchi}, {Skilton}, {Sol}, {Spangler},
  {Stawarz}, {Steenkamp}, {Stegmann}, {Superina}, {Szostek}, {Tam}, {Tavernet},
  {Terrier}, {Tibolla}, {van Eldik}, {Vasileiadis}, {Venter}, {Venter},
  {Vialle}, {Vincent}, {Vivier}, {V{\"o}lk}, {Volpe}, {Wagner}, {Ward},
  {Zdziarski}, \& {Zech}}]{Aharonian.2009}
{Aharonian}, F., {Akhperjanian}, A.~G., {Anton}, G., {et~al.} 2009, \aap, 503,
  817

\bibitem[{{Aharonian} \& {Neronov}(2005)}]{Aharonian.2005}
{Aharonian}, F. \& {Neronov}, A. 2005, \apj, 619, 306

\bibitem[{{Aharonian} {et~al.}(2019){Aharonian}, {Yang}, \& {de O{\~n}a
  Wilhelmi}}]{Aharonian.2019}
{Aharonian}, F., {Yang}, R., \& {de O{\~n}a Wilhelmi}, E. 2019, Nature
  Astronomy, 232

\bibitem[{{Aharonian}(2004)}]{Aharonian.2004}
{Aharonian}, F.~A. 2004, {Very high energy cosmic gamma radiation : a crucial
  window on the extreme Universe} (World Scientific Publishing Co)

\bibitem[{{Armillotta} {et~al.}(2019){Armillotta}, {Krumholz}, {Di Teodoro}, \&
  {McClure-Griffiths}}]{Armillotta.2019}
{Armillotta}, L., {Krumholz}, M.~R., {Di Teodoro}, E.~M., \&
  {McClure-Griffiths}, N.~M. 2019, MNRAS, 490, 4401

\bibitem[{{Barnes} {et~al.}(2017){Barnes}, {Longmore}, {Battersby}, {Bally},
  {Kruijssen}, {Henshaw}, \& {Walker}}]{Barnes.2017}
{Barnes}, A.~T., {Longmore}, S.~N., {Battersby}, C., {et~al.} 2017, \mnras,
  469, 2263

\bibitem[{{Battersby} {et~al.}(2020){Battersby}, {Keto}, {Walker}, {Barnes},
  {Callanan}, {Ginsburg}, {Hatchfield}, {Henshaw}, {Kauffmann}, {Kruijssen},
  {Longmore}, {Lu}, {Mills}, {Pillai}, {Zhang}, {Bally}, {Butterfield},
  {Contreras}, {Ho}, {Ott}, {Patel}, \& {Tolls}}]{Battersby.2020}
{Battersby}, C., {Keto}, E., {Walker}, D., {et~al.} 2020, \apjs, 249, 35

\bibitem[{{Blasi}(2013)}]{Blasi.2013}
{Blasi}, P. 2013, A\&A Reviews, 21, 70

\bibitem[{Bordoloi {et~al.}(2017)Bordoloi, Fox, Lockman, Wakker, Jenkins,
  Savage, Hernandez, Tumlinson, Bland-Hawthorn, \& Kim}]{Bordoloi.2017}
Bordoloi, R., Fox, A.~J., Lockman, F.~J., {et~al.} 2017, ApJ, 834, 191

\bibitem[{{Calder{\'o}n} {et~al.}(2020){Calder{\'o}n}, {Cuadra}, {Schartmann},
  {Burkert}, \& {Russell}}]{Calderon.2020}
{Calder{\'o}n}, D., {Cuadra}, J., {Schartmann}, M., {Burkert}, A., \&
  {Russell}, C. M.~P. 2020, \apj, 888, L2

\bibitem[{{Chernyakova} {et~al.}(2011){Chernyakova}, {Malyshev}, {Aharonian},
  {Crocker}, \& {Jones}}]{Chernyakova.2011}
{Chernyakova}, M., {Malyshev}, D., {Aharonian}, F.~A., {Crocker}, R.~M., \&
  {Jones}, D.~I. 2011, \apj, 726, 60

\bibitem[{{Clarkson} {et~al.}(2012){Clarkson}, {Ghez}, {Morris}, {Lu},
  {Stolte}, {McCrady}, {Do}, \& {Yelda}}]{Clarkson.2012}
{Clarkson}, W.~I., {Ghez}, A.~M., {Morris}, M.~R., {et~al.} 2012, \apj, 751,
  132

\bibitem[{{Crocker} {et~al.}(2010){Crocker}, {Jones}, {Melia}, {Ott}, \&
  {Protheroe}}]{Crocker.2010}
{Crocker}, R.~M., {Jones}, D.~I., {Melia}, F., {Ott}, J., \& {Protheroe}, R.~J.
  2010, Nature, 463, 65

\bibitem[{{CTA Consortium}(2019)}]{CTA.2019}
{CTA Consortium}. 2019, {Science with the Cherenkov Telescope Array} (World
  Scientific Publishing Co)

\bibitem[{{D{\"o}rner} {et~al.}(2024){D{\"o}rner}, {Becker Tjus}, {Blomenkamp},
  {Fichtner}, {Franckowiak}, \& {Zaninger}}]{Dorner.2024}
{D{\"o}rner}, J., {Becker Tjus}, J., {Blomenkamp}, P.~S., {et~al.} 2024, \apj,
  965, 180

\bibitem[{{Ekers} {et~al.}(1983){Ekers}, {van Gorkom}, {Schwarz}, \&
  {Goss}}]{Ekers.1983}
{Ekers}, R.~D., {van Gorkom}, J.~H., {Schwarz}, U.~J., \& {Goss}, W.~M. 1983,
  A\&A, 122, 143

\bibitem[{{Fatuzzo} {et~al.}(2006){Fatuzzo}, {Adams}, \&
  {Melia}}]{Fatuzzo.2006}
{Fatuzzo}, M., {Adams}, F.~C., \& {Melia}, F. 2006, \apjl, 653, L49

\bibitem[{{Fryer} {et~al.}(2006){Fryer}, {Rockefeller}, {Hungerford}, \&
  {Melia}}]{Fryer.2006}
{Fryer}, C.~L., {Rockefeller}, G., {Hungerford}, A., \& {Melia}, F. 2006, \apj,
  638, 786

\bibitem[{{Gaggero} {et~al.}(2017){Gaggero}, {Grasso}, {Marinelli}, {Taoso}, \&
  {Urbano}}]{Gaggero.2017}
{Gaggero}, D., {Grasso}, D., {Marinelli}, A., {Taoso}, M., \& {Urbano}, A.
  2017, Phys. Rev. Lett., 119, 031101

\bibitem[{{Gaisser} {et~al.}(2016){Gaisser}, {Engel}, \&
  {Resconi}}]{Gaisser.2016}
{Gaisser}, T.~K., {Engel}, R., \& {Resconi}, E. 2016, {Cosmic Rays and Particle
  Physics} (Cambridge University Press)

\bibitem[{{Genzel} {et~al.}(2010){Genzel}, {Eisenhauer}, \&
  {Gillessen}}]{Genzel.2010}
{Genzel}, R., {Eisenhauer}, F., \& {Gillessen}, S. 2010, Reviews of Modern
  Physics, 82, 3121

\bibitem[{{Ginsburg} \& {Kruijssen}(2018)}]{Ginsburg.2018}
{Ginsburg}, A. \& {Kruijssen}, J.~M.~D. 2018, \apjl, 864, L17

\bibitem[{{Guenduez} {et~al.}(2020){Guenduez}, {Becker Tjus}, {Ferri{\`e}re},
  \& {Dettmar}}]{Guenduez.2020}
{Guenduez}, M., {Becker Tjus}, J., {Ferri{\`e}re}, K., \& {Dettmar}, R.~J.
  2020, \aap, 644, A71

\bibitem[{{Guo} \& {Mathews}(2012)}]{Guo.2012}
{Guo}, F. \& {Mathews}, W.~G. 2012, \apj, 756, 181

\bibitem[{{Henshaw} {et~al.}(2023){Henshaw}, {Barnes}, {Battersby}, {Ginsburg},
  {Sormani}, \& {Walker}}]{Henshaw.2023}
{Henshaw}, J.~D., {Barnes}, A.~T., {Battersby}, C., {et~al.} 2023, in
  Astronomical Society of the Pacific Conference Series, Vol. 534, Protostars
  and Planets VII, ed. S.~{Inutsuka}, Y.~{Aikawa}, T.~{Muto}, K.~{Tomida}, \&
  M.~{Tamura}, 83

\bibitem[{{H.E.S.S. Collaboration} {et~al.}(2018){H.E.S.S. Collaboration},
  {Abdalla}, {Abramowski}, {Aharonian}, {Ait Benkhali}, {Akhperjanian},
  {Andersson}, {Ang{\"u}ner}, {Arakawa}, {Arrieta}, \& et~al.}]{HESS.2018}
{H.E.S.S. Collaboration}, {Abdalla}, H., {Abramowski}, A., {et~al.} 2018, A\&A,
  612, A9

\bibitem[{{H.E.S.S. Collaboration} {et~al.}(2016){H.E.S.S. Collaboration},
  {Abramowski}, {Aharonian}, {Benkhali}, {Akhperjanian}, {Ang{\"u}ner},
  {Backes}, {Balzer}, {Becherini}, {Tjus}, \& et~al.}]{HESS.2016}
{H.E.S.S. Collaboration}, {Abramowski}, A., {Aharonian}, F., {et~al.} 2016,
  Nature, 531, 476

\bibitem[{{Huang} {et~al.}(2021){Huang}, {Yuan}, \& {Fan}}]{Huang.2021}
{Huang}, X., {Yuan}, Q., \& {Fan}, Y.-Z. 2021, Nature Communications, 12, 6169

\bibitem[{{Kohno} {et~al.}(2004){Kohno}, {Yamamoto}, {Kawabe}, {Ezawa},
  {Sakamoto}, {Ukita}, {Hasegawa}, {Matsuo}, {Tatematsu}, {Sekimoto}, {Sunada},
  {Saito}, {Iwashita}, {Takahashi}, {Nakanishi}, {Yamaguchi}, {Kamazaki},
  {Sekiguchi}, {Ariyoshi}, {Yokogawa}, {Sugimoto}, {Toba}, {Oka}, {Sakai},
  {Tanaka}, {Takahashi}, {Hayakawa}, {Okuda}, {Muraoka}, {Fukui}, {Onishi},
  {Mizuno}, {Moriguchi}, {Minamidani}, {Mizuno}, {Suzuki}, {Ogawa}, {Yonekura},
  {Asayama}, {Kimura}, {Bronfman}, \& {Aste Team}}]{Kohno.2004}
{Kohno}, K., {Yamamoto}, S., {Kawabe}, R., {et~al.} 2004, in The Dense
  Interstellar Medium in Galaxies, Vol.~91 (Springer Berlin Heidelberg), 349

\bibitem[{{Kruijssen} {et~al.}(2015){Kruijssen}, {Dale}, \&
  {Longmore}}]{Kruijssen.2015}
{Kruijssen}, J.~M.~D., {Dale}, J.~E., \& {Longmore}, S.~N. 2015, MNRAS, 447,
  1059

\bibitem[{{Launhardt} {et~al.}(2002){Launhardt}, {Zylka}, \&
  {Mezger}}]{Launhardt.2002}
{Launhardt}, R., {Zylka}, R., \& {Mezger}, P.~G. 2002, \aap, 384, 112

\bibitem[{{Longair}(2011)}]{Longair.2011}
{Longair}, M.~S. 2011, {High Energy Astrophysics} (Cambridge University Press)

\bibitem[{{Maeda} {et~al.}(2002){Maeda}, {Baganoff}, {Feigelson}, {Morris},
  {Bautz}, {Brandt}, {Burrows}, {Doty}, {Garmire}, {Pravdo}, {Ricker}, \&
  {Townsley}}]{Maeda.2002}
{Maeda}, Y., {Baganoff}, F.~K., {Feigelson}, E.~D., {et~al.} 2002, ApJ, 570,
  671

\bibitem[{{MAGIC Collaboration} {et~al.}(2020){MAGIC Collaboration}, {Acciari},
  {Ansoldi}, {Antonelli}, {Arbet Engels}, {Baack}, {Babi{\'c}}, {Banerjee},
  {Barres de Almeida}, {Barrio}, {Becerra Gonz{\'a}lez}, {Bednarek},
  {Bellizzi}, {Bernardini}, {Berti}, {Besenrieder}, {Bhattacharyya},
  {Bigongiari}, {Biland}, {Blanch}, {Bonnoli}, {Bo{\v{s}}njak}, {Busetto},
  {Carosi}, {Ceribella}, {Chai}, {Chilingaryan}, {Cikota}, {Colak}, {Colin},
  {Colombo}, {Contreras}, {Cortina}, {Covino}, {D'Elia}, {da Vela}, {Dazzi},
  {de Angelis}, {de Lotto}, {Delfino}, {Delgado}, {Depaoli}, {di Pierro}, {di
  Venere}, {Do Souto Espi{\~n}eira}, {Dominis Prester}, {Donini}, {Dorner},
  {Doro}, {Elsaesser}, {Fallah Ramazani}, {Fattorini}, {Fern{\'a}ndez-Barral},
  {Ferrara}, {Fidalgo}, {Foffano}, {Fonseca}, {Font}, {Fruck}, {Fukami},
  {Garc{\'\i}a L{\'o}pez}, {Garczarczyk}, {Gasparyan}, {Gaug}, {Giglietto},
  {Giordano}, {Godinovi{\'c}}, {Green}, {Guberman}, {Hadasch}, {Hahn},
  {Herrera}, {Hoang}, {Hrupec}, {H{\"u}tten}, {Inada}, {Inoue}, {Ishio},
  {Iwamura}, {Jouvin}, {Kerszberg}, {Kubo}, {Kushida}, {Lamastra}, {Lelas},
  {Leone}, {Lindfors}, {Lombardi}, {Longo}, {L{\'o}pez}, {L{\'o}pez-Coto},
  {L{\'o}pez-Oramas}, {Loporchio}, {Machado de Oliveira Fraga}, {Maggio},
  {Majumdar}, {Makariev}, {Mallamaci}, {Maneva}, {Manganaro}, {Mannheim},
  {Maraschi}, {Mariotti}, {Mart{\'\i}nez}, {Masuda}, {Mazin},
  {Mi{\'c}anovi{\'c}}, {Miceli}, {Minev}, {Miranda}, {Mirzoyan}, {Molina},
  {Moralejo}, {Morcuende}, {Moreno}, {Moretti}, {Munar-Adrover}, {Neustroev},
  {Nigro}, {Nilsson}, {Ninci}, {Nishijima}, {Noda}, {Nogu{\'e}s}, {N{\"o}the},
  {Nozaki}, {Paiano}, {Palacio}, {Palatiello}, {Paneque}, {Paoletti},
  {Paredes}, {Pe{\~n}il}, {Peresano}, {Persic}, {Prada Moroni}, {Prandini},
  {Puljak}, {Rhode}, {Rib{\'o}}, {Rico}, {Righi}, {Rugliancich}, {Saha},
  {Sahakyan}, {Saito}, {Sakurai}, {Satalecka}, {Schmidt}, {Schweizer},
  {Sitarek}, {{\v{S}}nidari{\'c}}, {Sobczynska}, {Somero}, {Stamerra}, {Strom},
  {Strzys}, {Suda}, {Suri{\'c}}, {Takahashi}, {Tavecchio}, {Temnikov},
  {Terzi{\'c}}, {Teshima}, {Torres-Alb{\`a}}, {Tosti}, {Tsujimoto}, {Vagelli},
  {van Scherpenberg}, {Vanzo}, {Vazquez Acosta}, {Vigorito}, {Vitale}, {Vovk},
  {Will}, \& {Zari{\'c}}}]{MAGIC.2020}
{MAGIC Collaboration}, {Acciari}, V.~A., {Ansoldi}, S., {et~al.} 2020, \aap,
  642, A190

\bibitem[{{Muena} {et~al.}(2024){Muena}, {Riquelme}, {Reisenegger}, \&
  {Sandoval}}]{Muena.2024}
{Muena}, C., {Riquelme}, M., {Reisenegger}, A., \& {Sandoval}, A. 2024, \aap,
  689, A216

\bibitem[{{Oka} {et~al.}(2012){Oka}, {Onodera}, {Nagai}, {Tanaka}, {Matsumura},
  \& {Kamegai}}]{Oka.2012}
{Oka}, T., {Onodera}, Y., {Nagai}, M., {et~al.} 2012, \apjs, 201, 14

\bibitem[{{Ridley} {et~al.}(2017){Ridley}, {Sormani}, {Tre{\ss}}, {Magorrian},
  \& {Klessen}}]{Ridley.2017}
{Ridley}, M. G.~L., {Sormani}, M.~C., {Tre{\ss}}, R.~G., {Magorrian}, J., \&
  {Klessen}, R.~S. 2017, MNRAS, 469, 2251

\bibitem[{{Rockefeller} {et~al.}(2005){Rockefeller}, {Fryer}, {Baganoff}, \&
  {Melia}}]{Rockefeller.2005}
{Rockefeller}, G., {Fryer}, C.~L., {Baganoff}, F.~K., \& {Melia}, F. 2005,
  \apj, 635, L141

\bibitem[{{Sawada} {et~al.}(2004){Sawada}, {Hasegawa}, {Handa}, \&
  {Cohen}}]{Sawada.2004}
{Sawada}, T., {Hasegawa}, T., {Handa}, T., \& {Cohen}, R.~J. 2004, MNRAS, 349,
  1167

\bibitem[{{Scherer} {et~al.}(2022){Scherer}, {Cuadra}, \&
  {Bauer}}]{Scherer.2022}
{Scherer}, A., {Cuadra}, J., \& {Bauer}, F.~E. 2022, \aap, 659, A105, (Paper~I)

\bibitem[{{Scherer} {et~al.}(2023){Scherer}, {Cuadra}, \&
  {Bauer}}]{Scherer.2023}
{Scherer}, A., {Cuadra}, J., \& {Bauer}, F.~E. 2023, \aap, 679, A114,
  (Paper~II)

\bibitem[{{Sofue}(1995)}]{Sofue.1995}
{Sofue}, Y. 1995, PASJ, 47, 527

\bibitem[{{Sofue}(2022)}]{Sofue.2022}
{Sofue}, Y. 2022, \mnras, 516, 907

\bibitem[{{Strong} {et~al.}(2007){Strong}, {Moskalenko}, \&
  {Ptuskin}}]{Strong.2007}
{Strong}, A.~W., {Moskalenko}, I.~V., \& {Ptuskin}, V.~S. 2007, Annu. Rev.
  Nucl. Part. Sci., 57, 285

\bibitem[{{Su} {et~al.}(2010){Su}, {Slatyer}, \& {Finkbeiner}}]{Su.2010}
{Su}, M., {Slatyer}, T.~R., \& {Finkbeiner}, D.~P. 2010, ApJ, 724, 1044

\bibitem[{{Tsuboi} {et~al.}(1999){Tsuboi}, {Handa}, \& {Ukita}}]{Tsuboi.1999}
{Tsuboi}, M., {Handa}, T., \& {Ukita}, N. 1999, ApJ Suppl. Ser., 120, 1

\bibitem[{{Vieu} \& {Reville}(2023)}]{Vieu.2023}
{Vieu}, T. \& {Reville}, B. 2023, \mnras, 519, 136

\bibitem[{{Vieu} {et~al.}(2022){Vieu}, {Reville}, \& {Aharonian}}]{Vieu.2022}
{Vieu}, T., {Reville}, B., \& {Aharonian}, F. 2022, \mnras, 515, 2256

\bibitem[{{Yan} {et~al.}(2017){Yan}, {Walsh}, {Dawson}, {Macquart},
  {Blackwell}, {Burton}, {Rowell}, {Zhang}, {Xu}, {Tang}, \&
  {Hancock}}]{Yan.2017}
{Yan}, Q.-Z., {Walsh}, A.~J., {Dawson}, J.~R., {et~al.} 2017, \mnras, 471, 2523

\bibitem[{{Yang} {et~al.}(2015){Yang}, {Jones}, \& {Aharonian}}]{Yang.2015}
{Yang}, R.-z., {Jones}, D.~I., \& {Aharonian}, F. 2015, \aap, 580, A90

\end{thebibliography}

\begin{appendix}

\section{Model} \label{App_model}

In this appendix we summarise our physical and numerical model.  Further details are available in Papers I and II.

\subsection{Cosmic-ray model}

We modeled the CR dynamics using the standard diffusion equation for protons \citep{Aharonian.2004, Strong.2007}, neglecting reacceleration as its energy gain is much slower than regular acceleration in shock fronts \citep{Longair.2011,Blasi.2013},
\begin{equation}
\label{eq_dif}
\frac{\partial \psi(\vec{r},p,t)}{\partial t}=\vec{\nabla} \cdot (D(E_p) \cdot \vec{\nabla} \psi)-\vec{\nabla} \cdot (\vec{V} \psi)-\frac{\partial}{\partial p}\left(\left(\frac{ \mathrm{d}p}{\mathrm{d}t}\right)\psi\right)+Q(\vec{r},p,t),
\end{equation} 
where $\psi(\vec{r},p,t)$ is the CR density per unit of CR momentum $p$ at position $\vec{r}$ and time $t$, $D(E_p)$ is the diffusion coefficient, $E_p$ is the energy of ultra-relativistic particles, $\vec{V}$ is the advection velocity, and $Q(\vec{r},p,t)$ is the CR source term. In our current model, we consider only particles of $E_p=3\,$TeV; therefore, we transferred all energy of CRs between 0.1 and 3 TeV to these mono-energetic particles to be consistent with the CR energies selected in \citet{Gaggero.2017}, considering that $E_\gamma \approx 0.1E_p$, where $E_\gamma$ is the observed gamma-ray energy \citep{Aharonian.2004,Longair.2011}. In Papers I and II we used the same simplification for higher-energy particles because the analytical solutions of Eq. \ref{eq_dif} are already obtained with mono-energetic particles subject to standard diffusion.

\subsubsection{Sources} \label{subsub_source}

In Paper~I we considered the NSC and Sgr A East as CR accelerators in the GC, and in Paper~II, we added the AC and QC. We restricted our fiducial model to CR sources with well-constrained kinetic energies and a high certainty of being within the CMZ. More details are provided in Sect.~2.1.1 of Paper~II, which is summarized here.

For clusters of young massive stars with colliding stellar winds, we defined a continuous injection as $Q_{con}=A_0(E_p/E_0)^{-\Gamma}g(t)$, where $A_0$ is the CR injection rate, $E_0$ is the energy normalization, $\Gamma$ is the spectral index of the CR intrinsic spectrum, and $g(t)$ is a Heaviside function in the time interval $t_{age}<t$, where $t_{age}$ is the source age. The $A_0$ was chosen assuming that 1.6\% of the stellar winds’ kinetic power (computed from the mass-loss rate and typical stellar wind velocity) goes into acceleration of CRs between 0.1 and 3 TeV. This rate was found by integrating the proton energy spectrum $\propto E^{-2.3}$ \citep{HESS.2016,Aharonian.2019} and assuming that a canonical 10\% of their kinetic power is transferred to all CRs with energies above 1 GeV \citep{Blasi.2013,Aharonian.2004,Aharonian.2019}. The proposed continuous CR sources are: the NSC, modeled with a injection rate of $A_0=5.4 \times 10^{36}$ erg s$^{-1}$; the AC, modeled with $A_0=3.2 \times 10^{36}$ erg s$^{-1}$; and the QC, modeled with $A_0=1.8 \times 10^{36}$ erg s$^{-1}$. For each star cluster, we considered a $t_{age} = 10^6$ yr, since the stellar winds remain approximately constant over that timescale \citep{Calderon.2020,Vieu.2023}, the actual age of the young stellar populations in each cluster is of that order of magnitude \citep{Genzel.2010}, and an accurate assumption of $t_{age}$ makes no difference in this timescale within the CMZ.

For supernova explosions, we defined the impulsive injection rate as $Q_{imp}=B_0(E_p/E_0)^{-\Gamma} \delta(t-t_{age})$, where $B_0$ is the injection energy and $\delta(t)$ is a Dirac delta function (with units of s$^{-1}$). Again, we assumed that 1.6\% of the supernova explosion energy goes into the acceleration of CRs between 0.1 and 3 TeV. The proposed impulsive CR source is Sgr A East, which was modeled with an injection energy of $B_0=2.4\times 10^{49}$ erg and $t_{age} = 1700$ yr \citep{Rockefeller.2005,Fryer.2006}.

\subsubsection{Cosmic-ray dynamics} \label{Cr_dim}

We considered two different CR dynamics based on our previous works. More details are given in Sect.~2.1 of Paper~I and Sects.~2.1.2 and 2.1.3 of Paper~II, which are summarized here.

In high-density gas areas, a diffusion two orders of magnitude lower than that of the broader Galactic disk was constrained from gamma-ray observations \citep{Abeysekara.2017,Aharonian.2019}. Therefore, a low diffusion coefficient is appropriate for the CMZ. In Paper~I we simulated an isotropic diffusion quantified as $D_{GC}(E_p)= 10^{26}(E/10^{10}$ eV$)^\delta$ cm$^2$ s$^{-1}$ (two orders of magnitude lower than the standard Galactic interstellar medium) independent of the particle's location. In Paper~II we used the same low isotropic diffusion inside the CMZ, but a high isotropic diffusion was considered outside the CMZ, described by \hbox{$D_{ISM}(E_p)\sim 10^{28}(E_p/10^{10}$ eV$)^\delta$} cm$^2$ s$^{-1}$ \citep{Aharonian.2004,Strong.2007}. For both regimes, we considered $\delta=0.5$ for a Kraichnan spectrum for magnetic turbulence \citep{Aharonian.2004}.

Additionally, the outflow velocity of the Fermi bubbles \citep{Bordoloi.2017}, two $\sim$10 kpc-scale gamma-ray structures emanating from the GC \citep{Su.2010,Ackermann.2014}, suggests polar advective transport. Therefore, in Paper~II we included an additional ballistic motion with a velocity of 1000 km s$^{-1}$ toward either Galactic pole for all particles within $120\degr$ above or below Sgr A*. In Paper~I no advection was considered.

The diffusion coefficients for 3 TeV particles are $\approx 20$ times lower than those for 1 PeV particles, which were used in Papers I and II. Therefore, this slower diffusion increases the residence time of particles within the CMZ and decreases their diffusion radius with respect to our previous results. This effect also makes the transport by advection more important above and below Sgr A*.

\subsubsection{Energy loss and gamma-ray production} \label{subsub_ener_loss}

The production of $\pi^0$ in proton--proton collisions is by many orders of magnitude the dominant process of energy loss for CR protons \citep{Aharonian.2009, Longair.2011}. This process occurs when HE protons, or heavier nuclei, collide with protons in the ambient gas. These inelastic collisions produce secondary particles, such as $\pi^0$ (e.g., $p+p \rightarrow p+p+\pi^0$+...), which then decay into gamma rays \citep[$\pi^0 \rightarrow 2\gamma$;][]{Aharonian.2004,Longair.2011,Gaisser.2016}. The cooling time for that hadronic interaction is defined by $t_{pp}=5.3 \times 10^7 (n_\mathrm{H}/1~cm^{-3})^{-1}$ yr, where $n_H$ is the gas particle density \citep{Aharonian.2004}. For the CMZ, $n_\mathrm{H} \approx 300$ cm$^{-3}$ according to the modeled gas in Sect.~\ref{subsub-3D-CMZ}, and therefore $t_{pp} \approx 1.7 \times 10^5$ yr. Considering $D_{GC}$ for 3 TeV particles and the CMZ extension, the maximum time for protons to remain inside the CMZ could be higher than $t_{pp}$. Therefore, we assumed that CRs lose all their kinetic energy after $1.7 \times 10^5$ yr within the CMZ. Finally, leptons do not produce gamma rays in this band due to the high magnetic field in the GC \citep{Crocker.2010}. 

\subsubsection{Numerical simulation} \label{subsub-num-sim}

We simulated mono-energetic particles of 3 TeV using the Monte Carlo method in a 3D domain to obtain the modeled CR energy density ($w_{CR}$), where $w_{CR} \propto \psi$. We performed eight CR simulations of $2.5\times 10^4$ test particles for each source, each CR dynamics, and each CMZ shape (disk or ring). We discretized the results on a 3D grid centered on Sgr A*, where $w_{CR}$ was computed in each bin. The grid covers Galactic longitudes $-1.07\degr<l<1.73\degr$, Galactic latitudes $-0.42\degr<b<0.40\degr$, and $\pm 132$ pc along the line of sight. We assumed a GC distance of 8.5 kpc, with cubic bins of size 12 pc $\times$ 12 pc $\times$ 12 pc, resulting in an angular resolution of $\approx$ 0.08$\degr$. Finally, for each model combination, all eigth simulations were averaged for every bin.

\subsection{3D CMZ distribution} \label{subsub-3D-CMZ}

We summarize here the 3D gas distributions chosen in Paper~I, which were used again in Paper~II and in this work (for more details, see Sect.~2.2 of Paper~I).  

We developed two models that distribute the observed CMZ gas along the line of sight, based on \citet{Kruijssen.2015} and \citet{Launhardt.2002}. We considered that the CMZ could be an elliptical ring or an elliptical disk, meaning with or without a large inner cavity, as constrained in \citet{Sofue.1995,Kruijssen.2015} and \citet{Sawada.2004,Yan.2017}, respectively. We Note that the latter option also includes models that do consider a cavity \citep{Ridley.2017,Armillotta.2019}, but one that is small or shallow enough that it can be neglected for our purposes. Within the disk or ring, we distributed uniformly the column density of molecular gas along the line-of-sight domain, to obtain an approximation of the 3D CMZ distribution. The total CMZ column density was measured from CO ($J=3-2$) lines observed by the Atacama Submillimeter Telescope Experiment \citep[ASTE;][]{Kohno.2004} and published by \citet{Oka.2012}.\footnote{Available at \url{https://www.nro.nao.ac.jp/~nro45mrt/html/results/data.html}} We used a molecular line for high-density areas to ensure we only traced the CMZ and not part of the Galactic disk. Finally, we computed $n_\mathrm{H}$ using the same 3D grid described in Sect.~\ref{subsub-num-sim}, considering the gas to be static.

\subsection{Gamma-ray synthetic maps}
\label{sub-gamma-maps}

As described in Sect.~\ref{subsub_ener_loss}, gamma rays are produced from proton--proton interactions. We computed the gamma-ray luminosity per unit volume ($L_\gamma$/$V_\mathrm{bin}$) in the same grid of Sect.~\ref{subsub-num-sim}, using the values of $w_{CR}$ and $n_\mathrm{H}$ obtained in each grid bin as \citep{Fatuzzo.2006,HESS.2016}
\begin{equation}
\label{eq_lum}
\frac{L_\gamma}{V_\mathrm{bin}} \approx \kappa_\mathrm{\pi}~\sigma_\mathrm{p-p}~c~\eta_\mathrm{N}~n_\mathrm{H}~w_\mathrm{bin},
\end{equation}
where $\kappa_\mathrm{\pi}$ is the fraction of kinetic energy of HE protons transferred to $\pi^0$ production, $\sigma_\mathrm{p-p}$ is the cross section for proton--proton interaction, and $\eta_\mathrm{N}$ is the gamma-ray contribution from heavier nuclei in CRs and ambient gas. For protons with energies in the  GeV--TeV range, $\kappa_\mathrm{\pi} \approx 0.18$ \citep{Fatuzzo.2006}; for mono-energetic particles of 3 TeV, $\sigma_\mathrm{p-p} \approx 43$ mb \citep{Aharonian.2004} and $\eta_\mathrm{N} \approx 1.5$ \citep{HESS.2016}. Next, we integrated $L_\gamma$/$V_\mathrm{bin}$ along each line of sight to obtain gamma-ray synthetic maps, where the gamma-ray luminosity per bin cross section was converted to a gamma-ray surface flux computed at the Earth reference frame. Our model neglects the gamma-ray foreground and background of the CMZ.

\section{Gamma-ray spectrum for Sgr A*}
\label{App_SgrA*_spec}

\begin{figure}[h]
\resizebox{\hsize}{!}{\includegraphics{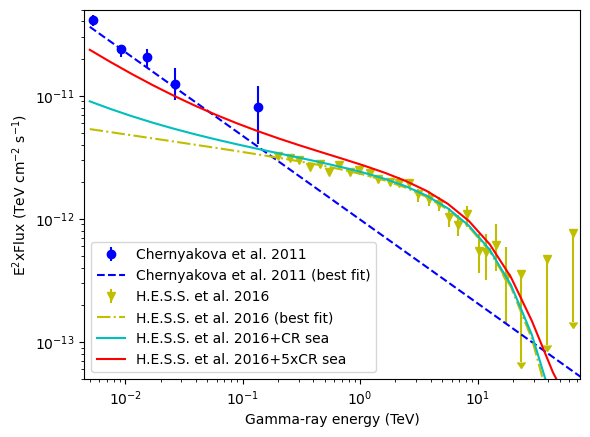}}
\caption{Same as Fig. \ref{fig_spec_cmz} but for Sgr A*. Here we additionally consider five times the CR sea observed in the Solar System.}
\label{fig_spec_sgrA}
\end{figure}

 In Fig. \ref{fig_spec_sgrA} is shown the HESS and Fermi LAT data of Sgr A* along with their best fits. For the HESS detection, we directly added the expected gamma-ray production by the CR sea measured in the Solar System and five times this contribution. 

\end{appendix}
\end{document}